\documentclass[journal]{IEEEtran}
\usepackage{subfigure}
\usepackage{graphicx}
\usepackage{amsmath}
\usepackage{multirow}
\usepackage[dvips]{epsfig}
\usepackage[latin1]{inputenc}
\usepackage[T1]{fontenc}
\usepackage{algorithm}
\usepackage{algorithmic}
\usepackage[dvips]{color}

\begin{document}

\title{Towards LP Modeling for Maximizing Throughput and Minimizing Routing Delay in Proactive Protocols in Wireless Multi-hop Networks}

\author{N. Javaid$^{\ddag}$, Z. A. Khan$^{\$}$, U. Qasim$^{\pounds}$, M. A. Khan$^{\ddag}$, K. Latif$^{\ddag}$, A. Javaid$^{\natural}$\\\vspace{0.4cm}
%    \IEEEauthorblockA{ \{nadeem.javaid,djouani@univ-paris12.fr\}}\\
        COMSATS Institute of IT, $^{\ddag}$Islamabad, $^{\natural}$Wah Cantt, Pakistan. \\
        $^{\$}$Faculty of Engineering, Dalhousie University, Halifax, Canada.\\
        $^{\pounds}$University of Alberta, Alberta, Canada.
     }

\maketitle

\begin{abstract}
Wireless Multi-hop Networks (WMhNs) provide users with the facility to communicate while moving with whatever the node speed, the node density and the number of traffic flows they want, without any unwanted delay and/or disruption. This paper contributes Linear Programming models (LP\_models) for WMhNs. In WMhNs, different routing protocols are used to facilitate users demand(s). To practically examine constraints of respective LP\_models over different routing techniques, we select three proactive routing protocols; Destination Sequence Distance Vector (DSDV), Fish-eye State Routing (FSR) and Optimized Link State Routing (OLSR). These protocols are simulated in two important scenarios regarding to user demands; mobilities and different network flows. To evaluate the performance, we further relate the protocols strategy effects on respective constraints in selected network scenarios.
\end{abstract}


\section{Background}

This work is devoted to study the routing capabilities of three proactive protocols named as Destination-Sequence Distance Vector (DSDV) \cite{1}, Fish-eye State Routing (FSR) \cite{2} and Optimized Link State Routing (OLSR) \cite{3} in different network cases of WMhNs.

In literature, we find different analysis on performance of routing protocols for different scenarios. A scalability analysis is presented in \cite{4}, which evaluates routing protocols with respect to different number of (CBR) resources. This analysis describes performance evaluation of AODV and DSR protocols influenced by the network size (up to $550$ nodes), nodes' mobility and density. The authors in \cite{5} evaluate the performance of DSR and AODV with varied number of sources  ($10$ to $40$ sources with different pause time). They demonstrate that even though DSR and AODV share a similar on-demand behavior, the differences in the protocol mechanics can lead to significant performance differentials. The problem from a different perspective is discussed in \cite{6}, using the simulation model with a dynamic network size and is examined practically for DSDV, AODV \cite{7}, DSR \cite{8} and Temporally Ordered Routing Algorithm (TORA).

The authors in \cite{9} examine the performance of proactive routing protocols. They set up a mathematical model to optimize proactive routing overhead without disturbing accuracy of routes. They present a generalized mathematical model for proactive routing protocol and specifically study the use of ACK mechanism. Finally they deduce that by optimizing the time interval of HELLO messages, proactive protocol will have less routing overhead and high delivery rate. Their evaluation based on mathematical model is generalized for proactive class, however, in our work, we specifically  discuss the behavior of reactive (AODV, DSR, DYMO) along with proactive protocols (DSDV, FSR and OLSR).

\section{Proactive Protocols with their basic Operations}

\subsection{DSDV}
DSDV protocol performs three type of maintenance operations; $LSM$, $RU\_per$ and $RU\_tri$ as mentioned in \cite{10}. Whereas, this protocol sends routing messages for $RU\_tri$ and $RU\_per$, because of link sensing from MAC layer. So, $CE$ of DSDV depends on the interval of $RU\_per$ and $RU\_tri$. Moreover, DSDV uses flooding mechanism to disseminate routing information. Let $CE_{RU\_per}^{DSDV}$ and $CE_{RU\_tri}^{DSDV}$ represents $CE$ of periodic and trigger updates of DSDV, and we can write the total $CE$ as:

\begin{eqnarray}
CE_{total}^{(DSDV)}=CE_{RU\_per}^{DSDV}+CE_{RU\_tri}^{DSDV}
\end{eqnarray}

\begin{equation*}
\begin{aligned}
&   CE_{RU\_per}^{(DSDV)}= \left(\frac{\tau_{NL}}{\tau_{RU\_per}}\right)\sum_{i=1}^{N}i       && \text{(1.a)}\\
&   EC_{RU\_tri}^{DSDV}= \int_{\tau_{NS}}^{\tau_{NL}} sgn|s_{lb}^{AR}| \sum_{i=1}^{N}i      && \text{(1.b)}
\end{aligned}
\end{equation*}
where, generation of $RU_{tri}$ depends on status of $lb$ among $AR$.

\subsection{FSR}
To avoid routing overhead, FSR only uses periodic maintenance operations; $LSM$ and $RU\_per$. For $LSM$ and $RU\_per$, MAC layer notification and Scope Routing \textit{(SR)} are performed, respectively. In \textit{SR}, diameter of whole network is divided into scopes and information is exchanged between scopes using graded-frequency technique. Two scopes; Inter-Scope and Intra-Scope are defined for FSR in \cite{2} and $CE$ for these scopes is given below:

\begin{eqnarray}
CE_{total}^{(FSR)}= CE_{RU\_per}^{IAS}+CE_{RU\_per}^{IES}
\end{eqnarray}

\begin{equation*}
\begin{aligned}
& EC_{RU\_per}^{IAS}= \left(\frac {\tau_{NL}}{\tau_{IAS}}\right) \sum_{i=1}^{N}\sum_{i=1}^{N_{IAS}}i  && \text{(2.a)}\\
& EC_{RU\_per}^{IES}= \left(\frac {\tau_{NL}}{\tau_{IES}}\right) \sum_{i=1}^{N}\sum_{i=1}^{N_{IES}}i   && \text{(2.b)}
\end{aligned}
\end{equation*}

Here, $\tau_{IAS}$ and $\tau_{IES}$ are IntraScope\_Interval and InterScope\_Interval, respectively (Table. 1). Whereas, ${N_{IAS}}$ and ${N_{IES}}$ represent total number of nodes in IntraScope (IAS) and InterScope (IES).

\subsection{OLSR}
In OLSR, $LSM$ and $RU\_tri$ are used to get information for links and routes. $LSM$ is performed by generating HELLO messages on routing layer after $HELLO\_INTERVAL$ ($LSM$ Table. 1). Whereas, $RU_{tri}$ is broadcasted through TC messages. The interval between successive $RU_{tri}$ depends on stability of MPRs. This stability is periodically confirmed through HELLO messages. On the other hand, to calculate topology information, TC messages are broadcasted. The broadcasting period of TC message depends on status of MPRs after $TC\_INTERVAL$ (default value as mentioned in Table 1) if MPRs are stable, while these messages are triggered and are transmitted to whole network in case of unstable MPRs, when node 6 detects link breakage then OLSR generates $RU_{tri}$.
The $CE$ of OLSR is given below:

\begin{eqnarray}
CE_{total}^{OLSR}=CE_{LSM}^{OLSR}+CE_{RU\_tri}^{OLSR}
\end{eqnarray}

\begin{equation*}
\begin{aligned}
\begin{split}
&EC_{LSM}^{OLSR}=\left(\frac{\tau_{NL}}{\tau_{HELLO}}\right)\sum_{i=1}^{N}nb_{i}
&& \text{(3.a)}\\
&EC_{RU\_tri}^{OLSR}=\begin{cases}
\displaystyle \int_{\tau_{NS}}^{\tau_{NL}}\sum_{i=1}^{N_{MPRs}} i\,\,If\,\,MPRs\,\,are\,\,stable\\
\displaystyle \int_{\tau_{NS}}^{\tau_{NL}}\sum_{i=1}^{N} i\,\,\, Otherwise
\end{cases}&& \text{(3.b)}
\end{split}
\end{aligned}
\end{equation*}

\begin{table}[ht]
\caption{Predefined Parameters Values}
\centering
\begin{tabular}{c c c }
\hline\hline
\textbf {Parameters} & \textbf {Used by (Protocol(s))} & \textbf {VALUES} \\ [0.5ex]
\hline
$RU_{per}$ Interval	& 	DSDV		&	15s 		\\
$LSM$ of MAC Interval	&	DSDV, FSR 	&	0.1s to 0.8s 	\\
HELLO\_INTERVAL		& 	OLSR 		& 	2s 		\\
TC\_INTERVAL (default)	&	OLSR 		& 	5s 		\\
TTL value for IntraScope	& 	FSR 		& 	2-hops 	\\
IntraScope\_Interval	& 	FSR 		& 	5s 		\\
TTL value for InterScope	& 	FSR 		& 	255-hops 	\\
InterScope\_Interval	& 	FSR 		& 	15 		\\ [1ex]
\hline
\end{tabular}
\label{table:nonlin}
\end{table}

\section{Simulations and Discussions}

To evaluate chosen protocols, we take different mobilities, scalabilities. We selected three performance metrics; throughput, $CT$ and $CE$. We analytically simulate $CT$ in terms of routing overhead and $CE$ in terms of frequency of topological exchange period. The performance metrics are measured through simulations in NS-2. For simulation setup, Random Way-Point is used as mobility models.  The area specified is $1000m \times 1000m$ field presenting a square space to allow mobile nodes to move inside. All of the nodes are provided with wireless links of a bandwidth of $2Mbps$ to transmit on. Simulations are run for $900s$ each. For evaluating mobilities effects, we vary pause time from $0s$ to $900s$ for $50$ nodes with speed $30m/s$. For evaluating different network flows with $15m/s$ speed and fixed pause time of $2s$, we vary nodes from $10$ nodes to $100$ nodes.

\subsection{Throughput}

\begin{figure} [h]
\includegraphics[height=50mm, width=80mm]{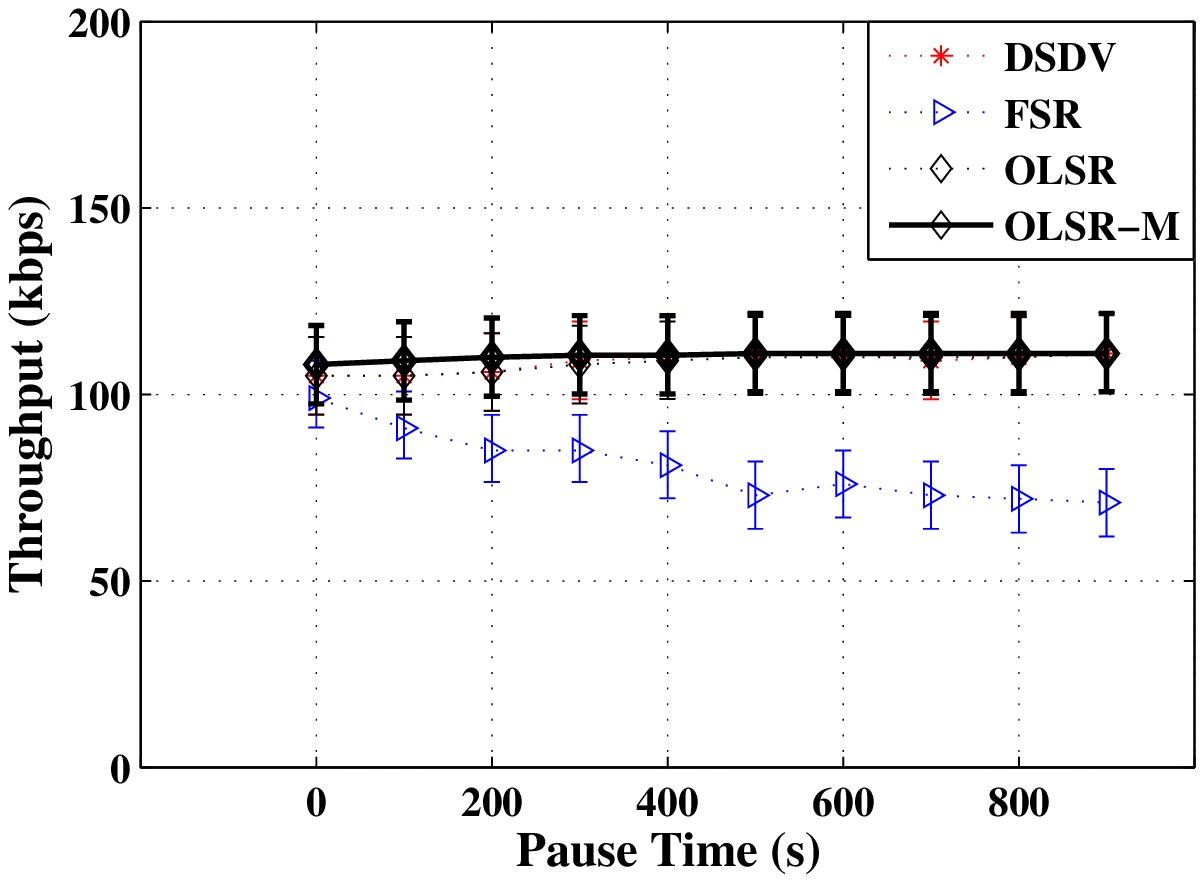}
\vspace{-0.4cm}
\caption{Throughput vs Pause Time of Proactive protocols}
\label{fig:05}
\end{figure}

Among proactive protocols, DSDV attains the highest throughput and shows efficient behavior in all pause times for, as shown in Fig. \ref{fig:05}. The reason for this good throughput is to use of route settling time; when the first data packet arrives, it is kept until the best route is found for a particular destination, thus overall satisfied constraints. Secondly, a decision may delay to advertise the routes which are about to change soon, thus damping fluctuations of the route tables. The rebroadcasts of the routes with the same sequence number are minimized by delaying the advertisement of unstabilized routes. This enhances the accuracy of valid routes resulting in the increased throughput of DSDV in all types of mobility rates, moreover, the updates are transmitting through NPDU's in small scalabilities. The reason for this gradual decrease with increasing mobility is the unavailability of valid routes due to its proactive nature. In static situation as well as low speed, in Fig. \ref{fig:05}, throughput is better as compared to moderate and relatively high mobility due to availability of stable entries for MPRs. Thus, in moderate and no mobilities OLSR performs well (Pause time more than $400s$ represents moderate mobilities, while pause time $900s$ means static mobile, because total simulation time is $900s$. Moreover, FSR does not trigger any control messages unlike DSDV  and OLSR when links breaks. Therefore, it is not as efficient as DSDV and OLSR.

\begin{figure} [h]
\includegraphics[height=50mm, width=80mm]{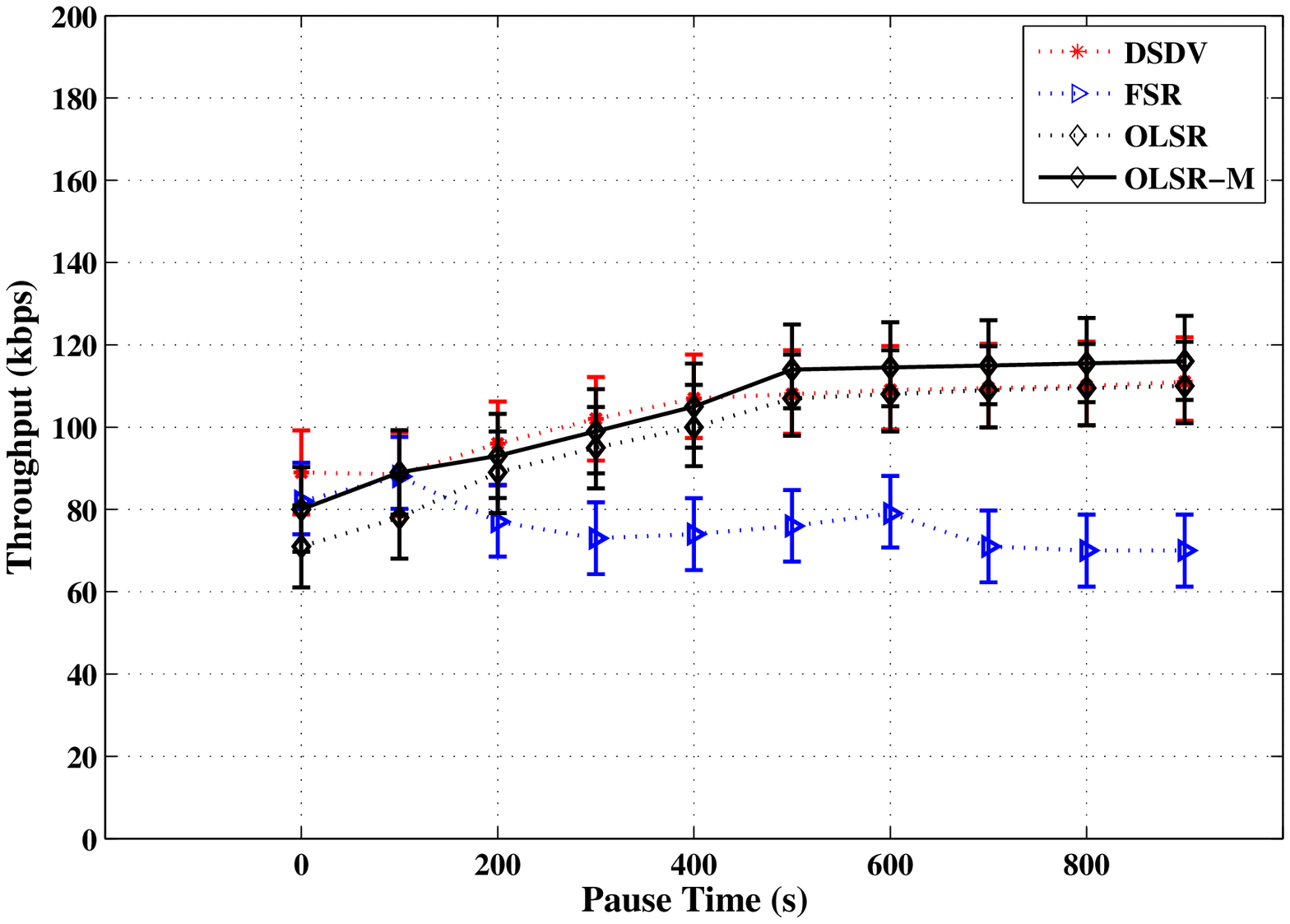}
\vspace{-0.4cm}
\caption{Throughput vs Pause Time of Proactive protocols}
\label{fig:06}
\end{figure}

\begin{figure} [h]
\includegraphics[height=50mm, width=80mm]{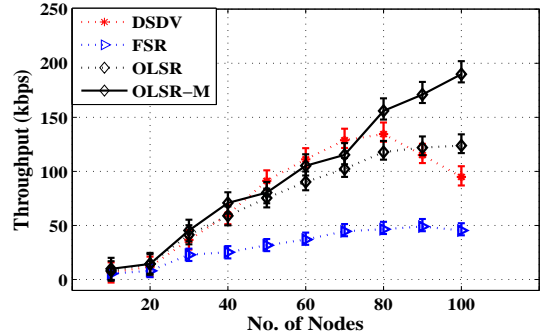}
\vspace{-0.4cm}
\caption{Throughput vs Scalability of Proactive protocols}
\label{fig:07}
\end{figure}

\begin{figure} [h]
\includegraphics[height=50mm, width=80mm]{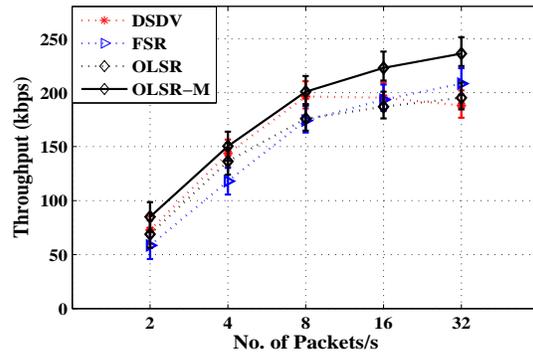}
\vspace{-0.4cm}
\caption{Throughput vs Transmission Rate of Proactive protocols}
\label{fig:08}
\end{figure}

FSR shows appreciable performance for varying traffic rates and OLSR is well scalable among proactive protocols. In medium and high traffic loads, FSR's performance is depicted in Fig. \ref{fig:07} and Fig. \ref{fig:08}. This is due to introduction of new technique of multi-level Fish-eye Scope (FS), that reduces routing overhead and works better when available bandwidth is low, thus increasing throughput in case of increased data traffic loads and reduces routing update overhead. Although, DSDV uses NPDUs to reduce routing transparency but $RU_{tri}$ causes routing overhead and degrades performance. OLSR uses MPRs for reduction of overhead but computation of these MPRs takes more bandwidth. Therefore its throughput is less than FSR. Further optimization helps FSR to only broadcast topology messages to neighbors in order to reduce flooding overhead. If FSR would have taken MAC layer feedback in case of link brakes then there might be exchange of messages to update neighbors, consuming bandwidth and lowering throughput. This faster discovery results in a better performance during high traffic loads. Simulation results of OLSR in Fig. \ref{fig:07} comparative to Fig. \ref{fig:08} show that it is scalable but less converged protocol for high traffic rates. This protocol is well suited for large and dense mobile networks, as it selects optimal routes (in terms of number of hops) and achieves more optimizations using MPRs. OLSR-M due to exchanging information of neighbors and with topology through frequent exchange results more throughput, as shown in Figs. \ref{fig:05}-\ref{fig:08}.

\subsection{Cost of Time}

\begin{figure} [bt]
\includegraphics[height=50mm, width=80mm]{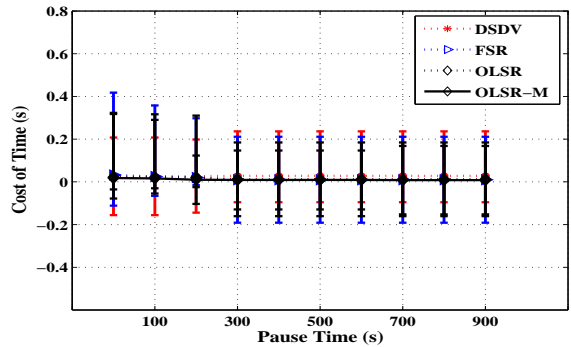}
\caption{Time Cost vs Pause Time of Proactive protocols}
\label{fig:09}
\end{figure}

\begin{figure} [bt]
\includegraphics[height=50mm, width=80mm]{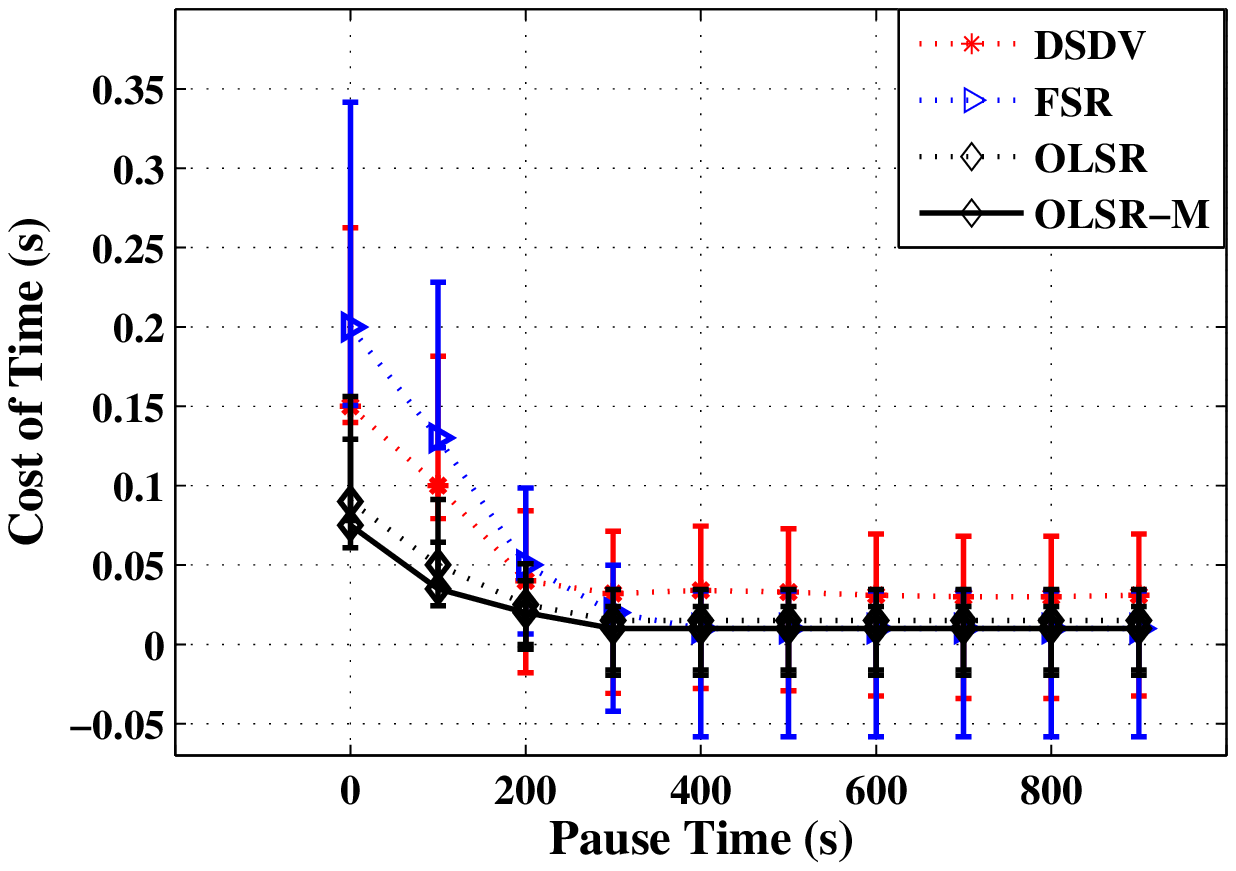}
\caption{Time Cost vs Pause Time of Proactive protocols}
\label{fig:10}
\end{figure}

In all proactive protocols, $CT$ value is directly proportional to speed and mobility, as depicted in Fig.~\ref{fig:09} and Fig.~\ref{fig:10}. DSDV possesses the highest delay cost among proactive in moderate and no mobility situations, as well as in all cases its E2ED is higher than OLSR. Because in DSDV, a data packet is kept for the duration between arrival of the first packet and selection of the best route for a particular destination. This selection creates delay in advertising routes which are about to change soon, thus causing damping fluctuations of the route tables. Furthermore, advertisement of the routes which are not stabilized yet is delayed in order to reduce the number of rebroadcasts of possible route entries that normally arrive with the same sequence number. FSR at higher mobilities produces the highest $CT$ value among proactive protocols. Due to graded-frequency mechanism when mobility increases, routes to remote destinations become less accurate. However, when a packet approaches its destination, it finds increasingly accurate routing instructions as it enters sectors with a higher refresh rate. At moderate and no mobilities at all speeds, the value of end to end delay is the same as well as this delay is less than other proactive protocol due to SR.

%E2ED Figures

\begin{figure} [bt]
\includegraphics[height=50mm, width=80mm]{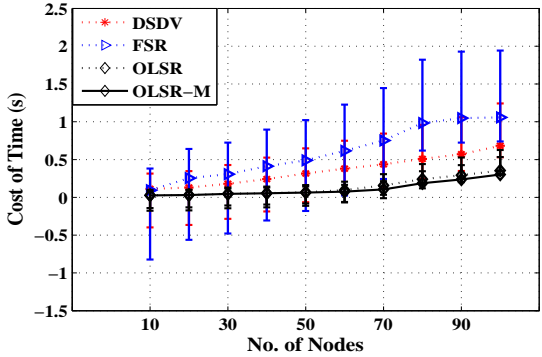}
\caption{Time Cost vs Scalability of Proactive protocols}
\label{fig:11}
\end{figure}

\begin{figure} [bt]
\includegraphics[height=50mm, width=80mm]{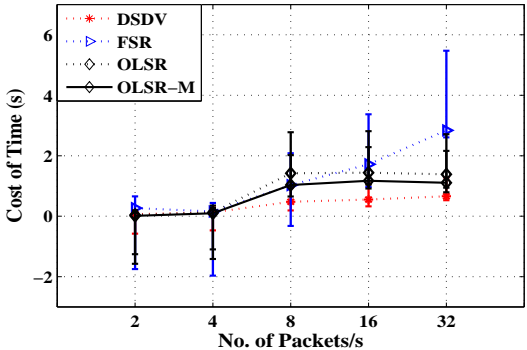}
\caption{Time Cost vs Transmission Rate of Proactive protocols}
\label{fig:12}
\end{figure}

FSR overall suffers higher delay in scalabilities due to retain route entries for each destination, this protocol maintains low single packet latency when population is small as shown in Fig.~\ref{fig:11} and Fig.~\ref{fig:12}. The graded-frequency mechanism is used to find destination to keep routing overhead low. FSR exchanges updates more frequently to the near destinations. Thus, in higher data rates or more scalabilities this protocol attains more $CT$ value. The reason for delay in DSDV is that it waits to transmit a data packet for an interval between arrival of first route and the best route. This selection creates delay in advertising routes which are about to change soon. A node uses new entry for subsequent forwarding decisions and route settling time is used to decide how long to wait before advertising it. This strategy helps to compute accurate route but produces more delay. Small values of $CT$ for OLSR are seen among proactive protocols in all scalabilities, because, MPRs provides efficient flooding control mechanism; instead of broadcasting, control packets are exchanged with neighbors only. In OLSR-M, routing latency is further decreased as compared to OLSR due decreasing $RU_{Tri}$ and $LSM_{Per}$ intervals (Figs.~\ref{fig:09}-\ref{fig:12}).

\vspace{10pt}
\section{Conclusion}

To practically examine constraints of respective LP\_models over proactive routing protocols, we select DSDV, FSR and OLSR. We relate the effects of routing strategies of respective protocols over WMhNs constraints to check energy efficient and delay reduction of these protocols in different scenarios in NS-2. DSDV shows more convergence to high dynamicties due to $RU_{tri}$ after detecting link layer feed back and provides optimal solution against constraints of $max\,\,\,T_{avg}$. FSR attains highest efficiency in more scalabilities by providing feasible solution against $max\,\,\,CE$ constraints due to scope routing. Whereas, OLSR and OLSR-M achieves highest throughput in scalabilities, because of feasible solution through MPRs against all constraints of $max\,\,\,CT$.

\end{document}